\documentclass{eas}
\usepackage{graphicx}
\begin{document}

\title{Astrophysical constraints on Dark Matter} 
\author{Charling Tao}\address{Centre de Physique des Particules de Marseille, IN2P3/CNRS 
and Universit\'e Aix-Marseille,  Marseille, France \\ 
Tsinghua Center for Astrophysics, Tsinghua University, Beijing, China}
\begin{abstract} Astrophysics gives evidence for the existence of Dark Matter 
and puts constraints on its nature. 
The Cold Dark Matter model has become "standard" cosmology
combined with a cosmological constant.  
There are indications that "Cold" Dark Matter could be "warmer"
than initially discussed.  
This paper reviews the main information on the Cold/Warm nature of Dark Matter.  

\end{abstract}
\maketitle
\section{Introduction}
Warm Dark Matter (WDM) has  become a hot topic and the debate is not settled. 
The Meudon Workshop 2011  was running in parallel to our CYGNUS Directional Direct Detection 
workshop followed in July by the 2011 Cosmology Colloque in Paris.  
Presentations from those two workshops are  available online.

Due to the restricted number of pages, I  skip the review on the astrophysical evidence for the existence of 
Dark Matter(DM). At all scales (galaxy, cluster and Universe) there are observations which point to the 
existence of a large quantity of DM. A small proportion of dark (ie, non luminous) matter may be baryonic:
astronomical bodies, such as black holes, massive compact halo objects, or cold molecular hydrogen, ...
DM, however, is currently defined in cosmological parameters fits, as non-baryonic and cold. 
Candidates include neutrinos, and  other hypothetical entities 
such as axions, or supersymmetry particles, ...

Non-baryonic DM  is classified as Hot Dark Matter (HDM), Warm Dark Matter (WDM), 
or Cold Dark Matter (CDM), depending on the velocity of the particles at decoupling.  
In the CDM hypothesis, the DM particles are massive and have small velocity. 
The most studied candidate is the supersymmetric neutralino of the MSSM 
(Minimal Supersymmetric Standard Model).

Hot DM particles (eg light neutrinos) would have velocities close to the speed of light.  
The DM  velocity has consequence on the large scale structure (LSS) formation.
The CDM model yields "bottom-up" hierarchical formation of structures in the universe
while HDM would induce preferentially  "top-down" formation.

In the nineties,  the development of N-body simulations of large scale structures (LSS)
led to a preference for CDM models over HDM.

\section{N-body simulations of LSS}

S. von Hoerner (\cite{Hoerner1, Hoerner2}) pioneered the work on N-body simulations in the early sixties.   
Many simulations use  only CDM, and thus include only the gravitational force. 
Incorporating baryons into the simulations dramatically increases their complexity and 
in the past, radical simplifications of the underlying physics was made.  Only in the last decade
have simulations tried to understand processes  that occur during galaxy formation.

N-body simulations of cosmological structures with CDM 
have shown that the radial profiles of the mass density and velocity dispersion of DM
haloes follow rather simple  universal functional forms
that are largely independent of halo properties such as mass,
environment, and formation history. 
For the density profile, there have been many discussions about the value of
the logarithmic slope of its central cusp, whether it is -1 (as for example in Hernquist (1990);
Navarro et al. (1997)) or -1.5 as in Moore et al. (1998). 

Results from more recent N-body simulations suggest actually a lack of
a definite inner slope: the density profile of the now better 
resolved DM haloes continues to 
flatten with decreasing
radius (e.g., Navarro et al. 2004; Merritt et al. 2005,
2006; Graham et al. 2006). Functional forms such as the Einasto (1969)  or the Prugniel
and Simien (1997) profiles, motivated by the Sersic profile for the surface
brightness of galaxies (Sersic 1968)  provide a
more accurate fit to the more recent simulations. 

Another parameter that has been used is the radial profile of 
the pseudo-phase-space density, $\rho/\sigma^3$, where $\sigma$ is either
the total velocity dispersion or the velocity dispersion in the radial 
direction. It seems well approximated by a power-law  (e.g.,
Taylor and Navarro 2001; Ascasibar et al. 2004; Dehnen and McLaughlin 2005; 
Hoffman et al. 2007, Stadel et al. 2009, ...). 

\section{Comparison of N-body simulations with observations}

At the end of the millenium, precision in N-body simulations of DM structures have shown some problems with a pure
CDM model at small scales: the predicted number of galactic satellites was not observed (cf. eg, Klypin et al.,1999)
and   a cusp/core controversy in galactic centers developed.
This has led some to conjecture WDM  to explain the discrepancy. 
However, new observed faint galactic satellites and other explanations 
for the observed galactic cores could allow the CDM model to survive. 
In the mean time, the size of  mini- voids in the local Universe
and  HI velocity functions and widths measurements  have increased the importance of 
the so-called "overabundance problem" in  pure $\Lambda$CDM simulations.
But the controversy is still on...

\subsection{Galaxy core profiles}

Observations of rotations curves  favour Burkert (1995) core profiles over the cusp profiles (cf eg., Donato et al. (2009), or Gentile et al.(2007))
from $\Lambda$CDM  N-body simulations (eg., Navarro,  Frenk and  White  (NFW 1996), or even Einasto form (1969)).  

It has been shown (eg, Mashchenko,  Couchman, and Wadsley,  2006) that 
stellar feedback can solve this difference by removing cusps. 
The numerical simulations with random bulk motions of gas (driven for example by supernovae explosions 
from star-forming galaxies)  can flatten  the central dark matter cusp on relatively short timescales ($\sim$108 years). 
Once removed, the cusp cannot be reintroduced during the subsequent mergers involved in the build-up 
of larger galaxies. As a consequence, in the present Universe both small and large galaxies would have 
flat dark matter core density profiles, in agreement with observations.   

Romano-Diaz et al. (2008)  proposed that baryons also can erase DM Cusps in Cosmological Galactic Halos.
They find a different evolution between the Pure DM (PDM) and Baryon+DM(BDM)
 models within the inner few 10 kpc region. The PDM model forms a  $R^{-1}$ cusp as expected, 
while the DM in the BDM model forms a larger isothermal 
 $R^{-2}$ cusp instead. The isothermal cusp is stable until $z \sim 1$  when it gradually levels off. 
This leveling proceeds from inside out and the final density slope is shallower than -1 within the central 3 kpc 
(i.e., expected size of the $R^{-1}$ cusp), tending to a flat core within $\sim$ 2 kpc. 
This effect cannot be explained by a finite resolution of the code, neither is it related to the energy feedback 
from stellar evolution or angular momentum transfer from the bar. 
Instead it can be associated with the action of DM+baryon subhalos heating up the cusp region 
via dynamical friction and forcing the DM in the cusp to flow out and to cool down. 

\subsection{Number of galactic satellites}
The recent discovery of many new DM dominated satellites of the Milky Way in the 
Sloan Digital Sky Survey (eg Belokurov, et al., 2010) has reduced the importance of 
the missing satellite issue. 

Maccio and Fontanot (2010) and Polisensky and Ricotti(2011) have given lower limits of a few keV to
DM particle mass from the number of Milky Way satellites,
since the number of satellites predicted decreases with decreasing mass of the DM particle. 
Assuming that the number of  satellites exceeds or equals the number of observed 
satellites of the Milky Way, Polisensky and Ricotti derive a lower limit on the DM particle mass of 13.3 keV (95$\%$CL)
for a sterile neutrino produced by the Dodelson and 
Widrow mechanism, 8.9 keV for the Shi and Fuller mechanism, 3.0 keV for the Higgs decay 
mechanism, and 2.3 keV for a thermal DM particle. 

These  lower limits are  comparable to constraints on WDM mass from
 Lyman-$\alpha$ forest modeling  (Narayanan et al. 2000; Viel et al. 2005,2008; Boyarsky et al. 2009a), high z quasar luminosity functions 
 (Song and Lee 2009), X-ray observations of the unresolved cosmic X-ray background,  DM halos from dwarf galaxy to cluster scales,  
(cf. eg. Boyanovsky, de Vega, Sanchez 2008 ; de Vega and Sanchez 2009 for reviews). 

\subsection{ HI velocity functions } In their very recent paper, 
Papastergis,  Martin, Giovanelli and Haynes  (2011) present results 
from 40$\%$ of the  ongoing wide-area, extragalactic HI-line, Arecibo Legacy Fast ALFA (ALFALFA) survey. 
They measure the space density of HI-bearing galaxies as a function of their observed velocity width 
(uncorrected for inclination) down to velocities of 20 km/s  and  confirm previous  indications
(Zavala et al., 2009;  Gottloeber, Hoffmann and Yepes, 2009; Trujillo-Gomez et al., 2010) of 
a substantial discrepancy at low widths between the observed distribution and the 
$\Lambda$CDM simulations. 

There is an overabundance of model galaxies by a factor of $\sim$ 10 compared
to observed dwarf galaxies with circular velocity $V_{circ} < $ 50 km/s. This is a serious problem for the
$\Lambda$CDM model: galaxies with these small circular velocities cannot be affected much by "normal"
physical processes (e.g., supernovae feedback or reionization of the Universe)
proposed for the solution of the satellite problem at $V_{circ}< $  30 km/s. 
The difference in abundance is a factor of about 8 at v= 50 km/s (which
corresponds to the resolution limit of the Zavala et al.(2009) simulation,
and implies a difference of a factor of $\sim$ 100 when extrapolated to the ALFALFA low-width limit (v = 20
km/s). 

Papastergis, Martin, Giovanelli and Haynes  (2011) also examine several solutions to the discrepancy: 
(i) a 1 keV WDM scenario and  (ii)  HI disks in low mass galaxies are usually not extended enough
 to probe the full amplitude of the galactic rotation curve. In this latter case, 
 they infer a relationship between the measured HI rotational 
velocity of a galaxy and the mass of its host CDM halo, which should be checked
to provide an important test of the validity of the established CDM model.

\subsection{Mini-voids size}
Tikhonov and Klypin (2010)  have studied the luminosity function,  peculiar velocities and  sizes of voids in the Local Volume 
 within the distance 4-8 Mpc.  The predictions of the standard cosmological $\Lambda$CDM model
give a factor of 10 more dwarf haloes as compared with the observed number of dwarf galaxies.
The theoretical void function matches the observations remarkably well only for haloes with circular velocities Vc 
larger than 40-45km/s.  For haloes with circular velocities $<$ 35km/s,
 there are too many small haloes in the $\Lambda$CDM model resulting in voids being too small, as compared with observations. 
The problem is that many of the observed dwarf galaxies have HI rotational velocities below 25 km/s
 that strictly contradicts the $\Lambda$CDM predictions. 
This is related to the  "overabundance problem", and could be solved by the same assumptions about
keV WDM or HI disks in low mass galaxies.

\subsection{Conclusion on the problems with $\Lambda$CDM?}
$\Lambda$CDM N-body simulations are fitting impressively well a wealth of data.
To-date discrepancies concern the "overabundance problem", which could, however, be due to the inability of 
HI to trace the maximum halo rotational velocity of low-mass systems.  So CDM is not dead yet!

The effects of WDM compared to CDM on the structure formation is 
to remove power from small scales, due to the large thermal velocities of the particles.
Future lensing projects like  EUCLID, can provide  measurements of a WDM mass for masses $<$ 2.5 keV,
since the  cosmic shear power spectra depends on the DM mass (Markovic et al. 2011), 
and departure from the CDM power spectra is not sensitive above roughly  2.5 keV.
In order to fully exploit future observations, models should be able to  predict the non-linear matter 
power spectrum at the level of 1 per cent or better for scales corresponding to comoving wavenumbers 
0.1 $<$ k $<$ 10 h Mpc$^{-1}$.
However baryonic and other astrophysics effects  (stellar, supernovae, AGN feedbacks, ... ) 
can have large impacts on the measured power spectrum at small scales. this has been verified by different groups of
N-body simulations: eg., Gottloeber et al., (2010), CLUES project;  Guillet, Teyssier and Colombi (2010), MareNostrum; 
Viel et al. (2011), Gadget-2;  van Daalen et al.(2011), OWLS, ... 

Comparison of N-body simulations of LSS with measurements can thus
exclude pure HDM solutions and set lower limits on the DM mass to be about a few keV.
Above a few keV, there is no current clues on how to separate WDM from CDM.

\section{Importance of Baryonic physics in N-body simulations}
The importance of baryonic physics in N-body simulations and weak lensing has been ignored till 2004.
White (2004) and Zhan and Knox (2004) calculated the effects of cooling  and intra-cluster gas on the lensing
power spectrum. These components  have each an effect of a few percent on the
lensing power spectrum at l around 3000, but with opposite signs. 
Jing et al. (2006) were the first to include in a N-body gas simulation  the physical processes
of radiative cooling and star formation, supernova feedback, outflows by galactic winds, and
a sub-resolution multiphase model for the interstellar medium.

More recently, Sales et al.(2010) studied the properties of simulated high-redshift galaxies
using cosmological N-body gas dynamical runs 
from the OverWhelmingly Large Simulations (OWLS) project. 
The different feedback models they use result in large variations in the 
abundance and structural properties of bright
 galaxies at z= 2. The OWLS simulations have also been used by 
 van Daalen, Schaye,  Booth, and  Dalla Vecchia (2011)  to study the distribution of power 
over different mass components, the back-reaction of the baryons on the CDM
 and the evolution of the dominant effects  on the matter power spectrum. 
Single baryonic processes are capable of changing the power spectrum by up 
to several tens of per cent. The simulation that includes AGN feedback, 
predicts a decrease in power relative to a dark matter only simulation ranging, at z= 0, 
from 1 per cent at k$\sim$ 0.3 h Mpc$^{-1}$ to 10 per cent at k$\sim$ 1 h Mpc$^{-1}$ and to 
30 per cent at k$\sim$ 10 h Mpc$^{-1}$. 
They confirm that baryons, and particularly AGN feedback, cannot be ignored in theoretical power 
spectra for k$>$ 0.3 h Mpc$^{-1}$. It is  necessary to improve our understanding of feedback 
processes in galaxy formation.

\section{Candidate DM: the sterile neutrino}
\subsection{The need for sterile neutrinos}
In the early times of the Standard Model of particles, neutrinos were thought to be 
massless and different lepton numbers were believed to be conserved. 
This was a reason for not introducing righthanded neutrinos. 
However, the observation of neutrino oscillations  in experiments with solar, atmospheric,
accelerator and reactor neutrinos requires the addition of new particles to the Standard Model.
Thus the interest in  "sterile" neutrinos which are right-handed, and have
very weak (if any) interactions, besides gravity...
Sterile neutrino  can be cold or warm DM depending on the models and parameters. 
Shaposhnikov, Boyarsky, and their collaborators presented many different models of sterile neutrinos.
A relatively new review of  astrophysical and cosmological constraints  on some models 
can be found in Boyarsky, Ruchayskiy and Shaposhnikov (2009).  The conclusion is that
 "Realistic Sterile Neutrino Dark Matter with KeV Mass does not Contradict Cosmological Bounds"
(Boyarsky,  Lesgourgues,  Ruchayskiy,  and Viel, 2009), in agreement with the many  astrophysical and laboratory
constraints on WDM mass (and neutrino mixing angles), 
which were first thoroughly investigated by Abazajian et al.(2001,2006,...) with the then existing data.

\subsection{Has a sterile neutrino of 5 KeV been found?}
Since 2006, several groups have searched for decaying DM 
(cf eg, the many papers of Boyarsky et al.., 2006 and after) and set
constraints on sterile neutrino model parameters. 
Loewenstein and Kusenko (2010) report the presence of a narrow emission
feature with energy 2.51$\pm$ 0.07(0.11) keV and flux [3.53 $\pm$ 1.95(2.77)] × $10^{-6}$ photons
 cm$^{-2} s^{-1}$ at 68$\%$
(90$\%$) confidence  in the Chandra X-ray Observatory spectrum of the ultra-faint dwarf spheroidal galaxy Willman 1. 
Interpreting this signal as an emission line from sterile neutrino radiative decay, the feature is
consistent with a sterile neutrino mass of 5.0 $\pm$  0.2 keV. 
But this signal is  too weak and would need confirmation before a claim of discovery can be made.

\section{Conclusion} Since Cygnus is a Directional Direct Detection workshop, 
news about a keV DM candidate discovery  can be a bit unsettling. 
It is important to keep an open eye on currently discussed candidates since it can have consequences on our projects.
If an Universe with only HDM can be excluded,  it is not possible to-date to rule out neither CDM nor WDM.
There are still many questions:
Can we trust present N-body simulations? They are impressive but halos from the simulations are not galaxies. 
Are all baryonic and other astrophysics effects well taken into account?     
The Wilman1 feature is not convincing, so whether DM is Cold or Warm and  in the keV range is still not settled. 
Furthermore, even if  WDM in the keV range existed, it is not excluded that more massive CDM would be also present.
Finally,  there are (many?) other particle candidates than keV sterile neutrinos. 
Lin et al. (2001) have, for example, proposed  the non-thermally  produced  decaying DM,   
which could reconcile  CDM and WDM. Some phenomenology has been presented by  Bi et al. (2010).
Considering the long timescales of Direct DM Detection efforts, 
I imagine  non-thermally produced decaying DM could be an alternative, welcome by this community...
 
\section{Ackowledgements:} Many thanks to Frederic Mayet, without whom I would not have reviewed the 
subject in Aussois. His gentle pressure has given birth to this written version... 
I am also grateful to Daniel Santos who  has kept, for so many years, the steady direction of TPCs 
for DM, and has continuously shared with me  the progress of his team. 
The workshop has allowed me to meet the new generation of enthusiastic 
DM research people and I enjoyed the Aussois environment.
The content of this talk has been enriched by discussions with  Zhang XinMin, Qin Bo, Shan HuanYuan, 
Bi XiaoJun and Zhan Hu.


\end{document}